# Accelerating Wilson Fermion Matrix Inversions by Means of the Stabilized Biconjugate Gradient Algorithm

A. Frommer, V. Hannemann and B. Nöckel
Mathematics Department
University of Wuppertal
D-42097 Wuppertal

Th. Lippert and K. Schilling
Physics Department
University of Wuppertal
D-42097 Wuppertal

**Abstract**

The stabilized biconjugate gradient algorithm BiCGStab recently presented by van der Vorst is applied to the inversion of the lattice fermion operator in the Wilson formulation of lattice Quantum Chromodynamics. Its computational efficiency is tested in a comparative study against the conjugate gradient and minimal residual methods. Both for quenched gauge configurations at $\beta = 6.0$ and gauge configurations with dynamical fermions at $\beta = 5.4$, we find BiCGStab to be superior to the other methods. BiCGStab turns out to be particularly useful in the chiral regime of small quark masses.

# 1 Introduction

Quantum Chromodynamics (QCD) is considered to be the fundamental theory of strong interactions among the quarks as constituents of matter. Its numerical evaluation by lattice techniques on a Euclidean space-time lattice [1] implies the computation of quark Greens functions in a chromodynamic background field. They are obtained by solving the inhomogeneous lattice-Dirac equation [2]

$$Mx = \phi. \tag{1}$$

The Dirac operator with its first-order derivatives in space and time translates into a matrix $M$ with nearest-neighbour interactions on a four-dimensional Euclidean space-time lattice [1], i.e. a sparse matrix that depends on the chromodynamic background gauge field.

There are two main areas in lattice-QCD, where the solution of Eq. 1 is of prime importance:

1. Hadron properties are computed from correlation functions built up from quark propagators. For the computation of quark propagators emanating from a fixed source in space-time, twelve columns of $M^{-1}$—corresponding to twelve point-like source terms $\phi$—have to be computed on a sufficiently large statistical sample of chromodynamic fields.

2. In the course of a simulation of QCD with dynamical fermions, the key algorithm is the so called Hybrid Monte Carlo method (HMC). In the HMC-algorithm, a deterministic molecular dynamics evolution of the system through phase space (hybrid molecular dynamics) [3] is followed by a Metropolis decision. Strictly speaking, the computation of the molecular dynamics force requires the solution of the equations

   $$M^\dagger M\, x = \phi \tag{2}$$

   during each individual step along the molecular dynamics history of the field system. In this context $\phi$ is a source term the elements of which are derived from independent Gaussian random numbers.

The common discretization schemes for the fermionic sector of lattice-QCD are Kogut-Susskind-fermions [4] and Wilson-fermions [5]. In recent years, the Wilson formulation of the fermionic sector of lattice-QCD has become more and more favoured as it suppresses fermion doubling [6] and appears to be more appropriate for the study of weak interactions.

The inversion problem is however rendered considerably more severe in the Wilson case, for three reasons: (a) the size of the Wilson-fermion matrix is four times larger than the corresponding Kogut-Susskind matrix, (b) the Wilson fermion matrix is a non-hermitian complex matrix,

$$M = \mathbf{1} - \kappa\, D, \tag{3}$$

and (c) its non-diagonal part $D$ is controlled by the hopping parameter $\kappa$, that needs tuning towards its critical value $\kappa_c$. Close to $\kappa_c$ is the critical regime of small eigenvalues, which in physical terms relates to the chiral regime of very small quark masses.

There are two common relaxation-type methods for solving Eq. 1: the minimal residual algorithm (MR) [7] applied directly to Eq. 1 and the conjugate gradient algorithm [8] applied to the normal equation

$$M^\dagger M\, x = M^\dagger \phi. \tag{4}$$

We will abbreviate this latter algorithm by CGNE (NE stands for *normal equation*) in order to distinguish it from the conjugate gradient method as applied to Eq. 1 directly, which will be denoted CG.

MR requires about half as many operations as CGNE and less storage. On the other hand, CGNE behaves much more stable numerically, especially when $M$ is approaching a singular matrix. This happens in the case when the hopping parameter $\kappa$ is tuned towards $\kappa_c$. For such values of $\kappa$, CGNE becomes definitely superior to MR, which shows a dramatic loss in efficiency near $\kappa_c$.

In the molecular dynamics application, Eq. 2, one would tend to employ CG, in view of the hermiticity of $M^\dagger M$. Nevertheless, one can as well apply a two-step solution scheme of type Eq. 5, based on MR,

$$M^\dagger\, y = \phi, \quad M\, x = y. \tag{5}$$

Here, both equations can be solved using a standard relaxation method such as MR. At $\kappa$ values far from $\kappa_c$ this might be superior to CG [9], but the more $\kappa_c$ is approached the more CG becomes advantageous. In both cases, however, the slow convergence rates—*i.e.* the phenomenon of critical slowing down of standard relaxation methods near $\kappa_c$—highly motivate the search for and the application of new algorithms.

An advance in this direction has been made by Plagge, who applied the biconjugate gradient (BiCG) method (see [10]) to the Higgs-Yukawa model. For certain choices of parameters, he demonstrated slight improvements in performance of BiCG, compared to CGNE and MR [11].

Recently, a new and very promising relaxation method has been proposed by van der Vorst [12]. It is derived from BiCG and is quoted as the stabilized biconjugate gradient (BiCGStab) algorithm. BiCGStab has been designed to treat non-hermitian matrices directly. It has been successfully applied to a large number of problems arising in various disciplines. In this paper, we will add to this list and demonstrate numerically, that BiCGStab turns out to be highly competitive in lattice QCD, in particular for the inversion of the non-hermitian complex Wilson fermion matrix. This statement holds, as we shall see, for both types of problems generically given by Eqs. 1 and 2.

In Section 2, we will recall the essential properties of BiCGStab and present its algorithmic formulation for complex, non-hermitian matrices. In Section 3.1, we shall benchmark CGNE, MR and BiCGStab as applied to the Wilson-fermion matrix, using quenched gauge configurations at $\beta = 6.0$ on a $16^4$ lattice. We will report results on convergence rates as well as real life implementation results on a 32-node Connection Machine CM5. Furthermore, we also employ BiCGStab as compared to CG and MR in a simulation with dynamical fermions, using the Hybrid Monte Carlo algorithm at $\beta = 5.4$ on a $8^4$ lattice in Section 3.2. We shall apply the new method to both, the full fermion determinant and its reformulation on an "even" subspace. We summarize our results in Section 4 and give an outlook as to preconditioning and work in progress.

# 2 BiCGStab and other cg-like methods

The *spectral condition number* $k(M)$ of a complex matrix $M$ is defined as the ratio of the largest to the smallest eigenvalue of $M^\dagger M$. (Since $M^\dagger M$ is hermitian positive definite whenever $M$ is nonsingular, all its eigenvalues are real and positive.) Thus, $k(M) \geq 1$ for all $M$ and $k(M) \rightarrow \infty$ as $M$ approaches singularity. The standard estimate (see [13], e.g.) for the CG efficiency on a hermitian positive definite matrix $A$ shows that the number of iterations required to achieve a given accuracy is roughly proportional to $\log(k(A))$. Note that $A = M^\dagger M$ for CGNE and $k(M^\dagger M) = (k(M))^2$. This explains the well known fact that CGNE is—albeit highly reliable—a rather slowly converging method.

In recent years, a lot of work in the numerical analysis community has gone into developing *cg-like* methods for solving Eq. 1 for general (i.e. non-hermitian, indefinite) matrices $M$. Reviews on this work can be found in [14, 15]. Contrary to CGNE, these new methods aim to avoid squaring the condition number while conserving several important characteristic properties of CG.

One of the most promising methods in this class is BiCGStab, the *stabilized biconjugate gradient* method proposed in [12]. In the present section we will outline the basic properties of BiCGStab without going into too many details. The convergence theory for cg-like methods is considerably less evident than for CG and there are several heuristics associated with it. We therefore restrict ourselves to exposing the major ideas. For easy reference, we present in Table 1 an overview of the arithmetic and the storage requirements of all methods discussed here. The first two columns of Table 1 contain the number of matrix-vector multiplications with $M$ and $M^\dagger$, followed by the number of inner products and of AXPY operations per iteration (an AXPY being an operation of the form $z = \alpha x + y$ with $x, y, z \in \mathbf{C}^n$, $\alpha \in \mathbf{C}$.). The final column gives the number of vectors in $\mathbf{C}^n$ which have to be stored in addition to $M$.

| Method | Mult. $M$ | Mult. $M^\dagger$ | inner prod. | AXPYs | storage |
|---|---|---|---|---|---|
| MR | 1 | – | 2 | 2 | 3 |
| CGNE | 1 | 1 | 2 | 3 | 4 |
| BiCG | 1 | 1 | 2 | 5 | 6 |
| CGS | 2 | – | 2 | 7 | 7 |
| BiCGStab | 2 | – | 4 | 6 | 7 |

Table 1: Arithmetic and storage requirements

BiCGStab may be regarded as a synthesis of two other cg-like methods,

the *biconjugate gradient* method BiCG (see [10]) and the *conjugate gradient squared* method CGS from [16]. All these methods proceed in a similar manner: During iteration $i$, given the current approximant $x_i$ as well as a search direction $p_i$ and the current residual $r_i = \phi - Mx_i$, the next iteration step $x_{i+1}$ is computed as $x_{i+1} = x_i - \alpha_i p_i,\ \alpha_i \in \mathbf{C}$. The new residual $r_{i+1}$ is updated from the previous value $r_i$ and a new search direction $p_{i+1}$ is generated. The methods basically differ in the way they generate $p_{i+1}$. A crucial requirement for the sake of computational efficiency is that the generation of $p_{i+1}$ should proceed via 'short recurrences' (involving only $r_i, p_i$ and one or two matrix-vector multiplications).

In BiCG the approximations $x_i$ are constructed such that $r_i$ is orthogonal to another additionally computed sequence of vectors $\hat{r}_0, \hat{r}_1, \ldots$ and vice versa. More precisely, we have $r_j^\dagger \hat{r}_i = 0,\ i = 0, 1, \ldots, j-1$ and $\hat{r}_j^\dagger r_i = 0,\ i = 0, 1, \ldots, j-1$. BiCG terminates after at most $n$ iterations, but for general $M$ it may break down due to division by zero before reaching the solution of Eq. 1. For theoretical purposes it is important to realize that the residuals of BiCG satisfy

$$r_j = P_j(M) r_0,\ \hat{r}_j = P_j(M^\dagger)\hat{r}_0$$

with polynomials $P_j$ of degree $\leq j$, the coefficients of which are implicitly computed during the iteration process. In the case of convergence, $P_j(M)r_0 \to 0$ and $P_j(M^\dagger)\hat{r}_0 \to 0$, so that we can regard $P_j(M)$ and $P_j(M^\dagger)$ as 'reductors' acting on $r_0$ and $\hat{r}_0$, respectively.

CGS is a modification of BiCG which avoids the computation of the $\hat{r}_j$. Instead, it produces residuals satisfying $r_j = P_j(M)^2 r_0$ with $P_j$ as above. So the reductor $P_j$ is applied twice in each step and, heuristically, we therefore expect this method to converge twice as fast as BiCG. Indeed, this behavior can be observed in practice. However, monitoring the size of the residuals $r_j$ of CGS sometimes reveals very large peaks. A simple explanation for this finding is that, although $P_j(M)$ is a reductor for $r_0$, it is not *eo ipso* one for $P_j(M)r_0$, and therefore $P_j(M)^2 r_0$ is not necessarily small. The large peaks can cause severe numerical problems in finite precision computations, due to cancellation errors. This phenomenon is particularly likely to occur if one starts the iteration close to the true solution [12].

Based on this observation, van der Vorst [12] developed the BiCGStab algorithm as a modification of CGS in which the residuals converge to zero in a smoother manner (provided no breakdown occurs). This algorithm again uses short recurrences only. The residuals now satisfy $r_j = Q_j(M)P_j(M)r_0$, where $Q_j$ is a polynomial of the form $Q_j(t) = \prod_{i=1}^{j}(1-\omega_i t)$ and the parameters $\omega_i$ are computed to render $Q_j(M)$ a good reductor for $P_j(M)r_0$. The full details of the resulting algorithm are given in Algorithm 1.

Algorithm 1: The BiCGStab algorithm

$r_0 = \phi - Mx_0$
$\hat{r}_0 = r_0$
$\rho_0 = \alpha = \omega_0 = 1$
$v_0 = p_0 = 0$
**for** $i = 1, 2, \cdots$
  $\rho_i = \hat{r}_0^\dagger r_{i-1}$
  $\beta = (\rho_i/\rho_{i-1})(\alpha/\omega_{i-1})$
  $p_i = r_{i-1} + \beta(p_{i-1} - \omega_{i-1} v_{i-1})$
  $v_i = Mp_i$
  $\alpha = \rho_i/\hat{r}_0^\dagger v_i$
  $s = r_{i-1} - \alpha v_i$
  $t = Ms$
  $\omega_i = t^\dagger s/t^\dagger t$
  $x_i = x_{i-1} + \omega_i s + \alpha p_i$
  $r_i = s - \omega_i t$
  `stop if` $r_i$ `is small enough`
**end for**

Though $\hat{r}_0$ can in principle be initialized arbitrarily in this algorithm, $\hat{r}_0 = r_0$ is the standard choice. If all eigenvalues of $M$ have positive real parts, it can be shown that there exists a choice for $\hat{r}_0$ (which is not known *a priori*), such that BiCGStab will converge towards the solution of Eq. 1 after at most $n$ steps. In particular, no breakdowns due to division by zero will occur.

# 3 Results

In the following, we will consider the application of BiCGStab to the computation of propagators and its use in lattice QCD with dynamical Wilson fermions.

The generic form of the the non-diagonal part of the Wilson fermion matrix $M$ in Eq. 3, reads [5]:

$$D(x,y) = \sum_{\mu=1}^{4} (\mathbf{r} - \gamma_\mu) U_\mu(x) \delta_{x,y-\mu} + (\mathbf{r} + \gamma_\mu) U_\mu(x-\mu) \delta_{x,y+\mu}. \tag{6}$$

We shall work with $\mathbf{r} = 1$. Due to the "$\gamma_5$"-symmetry of the matrix $M$,

$$\gamma_5 M^\dagger \gamma_5 = M, \tag{7}$$

the eigenvalues of $M$ come in complex-conjugate pairs. Hence the determinant of $M$ is real. $\kappa = \kappa_c$ is a singular point with $\det(M) = 0$, and for $\kappa < \kappa_c$, the matrix $M$ is positive definite, i.e. all its eigenvalues have positive real parts.

When building up $M$ one has the freedom of chosing any ordering scheme for the lattice sites. Subdividing the lattice in a checkerboard style into even and odd sites and numbering all even sites before the odd ones, one can cast $M$ into a block diagonal form [17]:

$$M = \begin{pmatrix} \mathbf{1} & -\kappa\, D_{eo} \\ -\kappa\, D_{oe} & \mathbf{1} \end{pmatrix}. \tag{8}$$

## 3.1 Computing propagators in quenched QCD

The computation of the propagator within a QCD background gauge field requires the solution of Eq. 1 for point-like or extended sources $\phi$. It has become standard to solve Eqs. 1 and 5 with $M$ given by Eq. 8 using "even-odd" preconditioning. To describe this preconditioning we rewrite Eq. 1, using Eq. 8, as

$$\begin{pmatrix} \mathbf{1} & -\kappa\, D_{eo} \\ -\kappa\, D_{oe} & \mathbf{1} \end{pmatrix} \begin{pmatrix} x_e \\ x_o \end{pmatrix} = \begin{pmatrix} \phi_e \\ \phi_o \end{pmatrix}. \tag{9}$$

The index 'e' denotes even, 'o' denotes odd sites. The equation separates into

$$M_e x_e = \tilde{\phi}_e \tag{10}$$

$$x_o = \phi_o + \kappa D_{oe} x_e, \tag{11}$$

where $M_e = \mathbf{1} - \kappa^2 D_{eo} D_{oe}$ and $\tilde{\phi}_e = \phi_e + \kappa D_{eo} \phi_o$.

Solving $Mx = \phi$ is thus essentially reduced to solving Eq. 10 for the even sites only. The odd sites can then be obtained by a simple substitution in Eq. 11. The matrix $M_e$ is second order in $\kappa$ [18] (first-order preconditioning) which translates into a factor of 2–3 in convergence rates. Although the dimensions of $M_e$ are only half those of $M$, the matrix $M_e$ is just as expensive to store as $M$, and one multiplication with $M_e$ requires approximately as many operations as with $M$. However, the additional storage requirements for the different iterative methods reported in Table 1 are now halved, since $M_e$ acts on the even sides only.

We have benchmarked three algorithms for solving Eq. 10—CGNE, MR, and BiCGStab—in a data-parallel implementation on a 32-node CM5 connection machine. For this purpose, we generated a sample of 10 decorrelated quenched $SU(3)$ gauge configurations, properly thermalized at $\beta = 6.0$. The propagators were computed on one source per configuration. For definiteness, we chose point-like sources, which are known to require more iterations than other sources.

The stopping criteria we chose were based on the accumulated computed residuals of each method. More precisely, we took

$$\frac{||M_e^\dagger M_e x_f - M_e^\dagger \tilde{\phi}_e||}{||M_e^\dagger x_f||} \leq r \quad \text{for CGNE} \tag{12}$$

and

$$\frac{||M_e x_f - \tilde{\phi}_e||}{||x_f||} \leq r \quad \text{for MR and BiCGStab.} \tag{13}$$

Here, $x_f$ denotes the final iterate, i.e. the first iterate satisfying Eq. 12 (for CGNE) or Eq. 13 (for MR and BiCGStab).

Of course, the criteria Eq. 12 and Eq. 13 are not equivalent, but we have $\|M_e^\dagger M_e x_f - M_e^\dagger \tilde{\phi}_e\| \leq \|M_e^\dagger\| \cdot \|M_e x_f - \tilde{\phi}_e\|$. Hence, the validity of Eq. 13 implies

$$\frac{||M_e^\dagger M_e x_f - M^\dagger \tilde{\phi}_e||}{||M_e^\dagger x_f||} \leq r \cdot \|M_e^\dagger\|.$$

Since the operator norm $\|M_e^\dagger\|$ is close to 1 this shows that Eq. 13 and Eq. 12 are at least comparable. In all our computations, at the end of *each* algorithm, we actually computed $\|M_e^\dagger M_e x_f - M_e^\dagger \tilde{\phi}_e\|/\|x_f\|$ *and* $\|M_e x_f - \tilde{\phi}_e\|/\|x_f\|$. Both these values turned out to be of equal order of magnitude in

all our examples, provided one choses the same $r$ in Eqs. 13 and 12. However, for BiCGStab and MR the value for $||M_e^\dagger M_e x_f - M_e^\dagger \tilde{\phi}_e|| / ||x_f||$ often was about a factor of two smaller than for CGNE so that Eq. 13 appears to be a more severe stopping criterion than Eq. 12. In this sense we might state that BiCGStab and MR produce more accurate solutions than CGNE. All our quenched computations have been done in double precision.

As a final cross check, we convinced ourselves that the resulting estimators from the different algorithms lie sufficiently close to each other, by looking at the norms,

$$\frac{||x_{f_{CGNE}} - x_{f_{BiCGStab}}||}{||x_{f_{CGNE}}||} \quad \text{and} \quad \frac{||x_{f_{CGNE}} - x_{f_{MR}}||}{||x_{f_{CGNE}}||}. \tag{14}$$

We want to demonstrate that BiCGStab is a superior tool in the realistic setting of QCD applications, i.e. in the appropriate quark mass range realized on a $16^4$ lattice. To set the stage, we illustrate the typical $\kappa$ dependence of the number of iteration steps, $N$, with $r = 10^{-12}$ in Eq. 12 and 13, respectively, on a given configuration in our sample. In Fig. 1, we normalized the measured $N_{CGNE}$ and $N_{MR}$ to $N_{BiCGStab}$. The mass scale has been computed from the relation

$$m_q = \frac{1}{2}\left(\frac{1}{\kappa} - \frac{1}{\kappa_c}\right). \tag{15}$$

We estimated $\kappa_c$ on this particular configuration from the convergence behaviour of $N_{CGNE}$ as function of the smallest eigenvalue [19]. In recent practice, most computational effort has been spent in the so-called chiral regime of small quark masses. As an example we quote the work of Guesken et al. [20], who performed quark propagator inversions in the mass range $0.1 > m_q \geq 0.029$ at $\beta = 6.0$. For the configuration considered above with $\kappa_c = 0.1568(2)$ this mass range translates into $0.152 < \kappa \leq 0.155$. In this interesting region, we find CGNE and MR to need more than twice as many iterations than BiCGStab.

The corresponding improvement factor in terms of real computer time on our 32-node CM5 is shown in Fig. 2. Throughout the $\kappa$-region, we find BiCGStab to beat the other methods. These results hold in particular for the chiral regime, where, compared to MR, we gain a factor of 1.2 - 1.5 using BiCGStab. Close to $\kappa_c$, MR exhibits the well-known loss in efficiency since it tends to stagnate near singularities due to roundoff errors. On the other side,

CGNE is slightly improving for $\kappa$ very close to $\kappa_c$ compared to BiCGStab. Yet BiCGStab beats CGNE by more than a factor of 2 at our smallest $\kappa$-value. Here, CGNE presumably reflects the finite size of the lattice, and we expect BiCGStab to behave even better for larger lattice sizes.

In Fig. 3, we display the iteration history for the three algorithms at $\kappa = 0.155$ on the same configuration as in Figs. 1 and 2. It shows that the convergence rate for the residual of MR equals that of CGNE. Though, at the beginning of the iterations, BiCGStab seems to follow CGNE closely, eventually its convergence rate turns out to be more than twice as fast. This illustrates the fact that—as opposed to CGNE and MR—BiCGStab avoids squaring the condition number. We observe a very smooth convergence for MR and with some restrictions also for CGNE. On the other hand, the residuals in BiCGStab exhibit its well-known peaked fluctuations. These fluctuations, however, are decreasing as the iteration progresses, and only at the very beginning they lead to residuals larger than those of MR and CGNE.

In a final cross check we compare the estimators $x_f$ produced by the different methods using the norms of Eqs. 14. We find values of $10^{-11}$. These results lie within the order of the chosen numerical precision.

We substantiate our observations by extension to our entire sample of ten statistically independent pure gauge configurations. The results are shown in Figs. 4, 5 and 6. The representation follows Fig. 1: The symbols denote the average values over the sample and the fluctuations are given by the extremal values from the individual configurations. They are visualized by bars. In Figs. 4 and 5, we present $N_{MR}$ and $N_{CGNE}$ normalized to $N_{BiCGStab}$ for two different values of $r$ in our stopping criteria, $r = 10^{-9}$ and $r = 10^{-12}$. As is to be expected from Fig. 3 already, there is no qualitative difference for both accuracies. The averages and fluctuations demonstrate that the conclusions gained on our individual example configuration appear to hold for the entire sample. The bandwidth of the fluctuations is very small and within the chiral region there is no overlap of both MR and CGNE results with BiCGStab results. The time ratios, as determined on the 32-node CM5, are given in Fig. 6 for $r = 10^{-12}$.

## 3.2 BiCGStab in Hybrid Monte Carlo simulations

The simulation of full QCD including dynamical Wilson fermions requires the efficient update of the square of the fermionic determinant, $\det(M)^2$. In the HMC algorithm, the starting point is the path integral representation of the determinant in terms of the pseudofermion fields $\phi$:

$$det(M)det(M^\dagger) = \int D\phi\, D\phi\ \exp[-\phi^*(M^\dagger M)^{-1}\phi]. \qquad (16)$$

The field $\phi$ does not participate in the evolution and is derived from Gaussian random numbers. The gauge variables $U$ are evolved through phase space in a fictitious computer time [21]. The moleculardynamics evolution repeatedly requires the solution of Eq. 2.

We have applied BiCGStab to two versions of the representation of the fermionic determinant, first the full determinant according to Eq. 16, and second its representation on an even subspace.

For both investigations, we worked on a $8^4$ lattice at $\beta = 5.4$. We performed HMC simulations at four values of $\kappa$ approaching $\kappa_c$. At each $\kappa$-value, we performed 400 thermalization steps. In view of the enormous expense in CPU-time, we only measured on ten configurations separated by 10 HMC sweeps each. Table 2 summarizes the HMC parameters used. We worked with single precision accuracy. Therefore, we expected larger numerical deviations in the accumulated vectors of the investigated relaxation methods. We shall discuss these deviations and again we shall show that the final estimates $x_f$ for the true solution $x$ from the different methods are sufficiently close.

| $\kappa$ | $\beta$ | # of MD-steps | dt | r |
|---|---|---|---|---|
| 0.158 | 5.4 | 25 | 0.02 | $10^{-8}$ |
| 0.160 | 5.4 | 25 | 0.02 | $10^{-8}$ |
| 0.161 | 5.4 | 25 | 0.02 | $10^{-8}$ |
| 0.162 | 5.4 | 25 | 0.02 | $10^{-8}$ |

Table 2: HMC parameters for both the full determinant and the determinant on an even subspace.

The $\kappa$-values chosen lie close to $\kappa_c = 0.1645$ as quoted in Ref. [22] for a $16^4$ lattice. Taking this value as an approximation on our $8^4$ system, $\kappa = 0.162$ would correspond to a quark mass $m_q \simeq 0.05$, according to Eq. 15. An $8^4$ lattice will of course exhibit severe finite size effects for such a small mass.

### 3.2.1 Full determinant

Here we are faced to solve a problem of type Eq. 2. So CG is applied directly to $M^\dagger M$. For MR and BiCGStab we perform the two-step procedure, Eq. 5, instead. This has the advantage that we could apply even-odd preconditioning as in Eq. 10. Though the two step procedure implies two separate iterations, even-odd preconditioning can yield a considerable improvement compared to the direct solution of Eq. 2.

In Fig. 7, we present the iteration numbers normalized again to $N_{BiCGStab}$. For BiCGStab and MR we summed up the numbers of iterations from both steps. Compared to CG, BiCGStab performs better by nearly a factor of 4 within the $\kappa$-range considered. Since in both procedures each iteration step requires two multiplications by $M$ and $M^\dagger$ or $M_e$ and $M_e^\dagger$ , resp., the improvement factor in terms of CPU-times is also quite close to 4. For MR, we roughly have to divide the iteration number ratio by a factor of two in order to get the time ratio. The resulting improvement of BiCGStab compared to MR therefore lies between 1.5 and 2. Again we encounter the well known loss of efficiency for MR approaching $\kappa_c$.

For each step of the two-step procedures with MR and BiCGStab, we imposed the same stopping criterion Eq. 13 with $r = 10^{-8}$. For CG we took Eq. 12, again with $r = 10^{-8}$. Note that in the two-step procedure we thus have

$$\begin{aligned}\|M_e^\dagger M_e x_e - M_e^\dagger \tilde{\phi}_e\| &= \|M_e^\dagger (M_e x_e - y_e) + M_e^\dagger y_e - M_e^\dagger \tilde{\phi}_e\| \\ &\leq \|M_e^\dagger\| \cdot \|M_e x_e - y_e\| + \|M_e^\dagger y_e - M_e^\dagger \tilde{\phi}_e\|,\end{aligned}$$

so that

$$\begin{aligned}\frac{\|M_e^\dagger M_e x_e - M_e^\dagger \tilde{\phi}_e\|}{\|x_e\|} &\leq \|M_e^\dagger\| \cdot \frac{\|M_e x_e - y_e\|}{\|x_e\|} + \frac{\|y_e\|}{\|x_e\|} \cdot \frac{\|M_e^\dagger y_e - M_e^\dagger \tilde{\phi}_e\|}{\|y_e\|}, \\ &\leq \|M_e^\dagger\| \cdot r + \frac{\|y_e\|}{\|x_e\|} \cdot r.\end{aligned}$$

Since $\|M_e^\dagger\|$ is small we can expect that the stopping criteria in the two-stage procedure will yield small residuals in the total equation, too. This is exactly

what we could observe in our practical experiments, where for all three methods we found only 10% deviation in the final residual. Note that this residual had to be recomputed directly at the end of each method, since due to the marked influence of rounding errors in these single precision computations the accumulated residuals were wrong by two orders of magnitude.

Furthermore, the differences in the norms of the final estimates, see Eq. 14, were verified to be less than $10^{-6}$.

### 3.2.2 Determinant on the even subspace

It has become standard to use the Wilson fermion determinant on an even subspace in order to accelerate the full QCD simulation [17]. The determinant can be reformulated:

$$\det(M) = \det\begin{pmatrix} \mathbf{1} & -\kappa\, D_{eo} \\ -\kappa\, D_{oe} & \mathbf{1} \end{pmatrix}. \tag{17}$$

The computation of the block determinant leads to

$$\det(M) = \det(\mathbf{1} - \kappa^2\, D_{eo} D_{oe}) = \det(M_e). \tag{18}$$

The simulation of the square of the determinant, Eq. 16, can be done on the even subspace sparing a factor of two in memory requirements and improving the condition of the numerical problem.

Using CG we have to multiply by the hermitian matrix $M_e^\dagger M_e$ during each iteration step. With MR and BiCGStab we again apply the two-step procedure as discussed before. We have employed the same HMC run parameters as in the simulation with the full determinant. In Fig. 8, we present the normalized iteration numbers. We find BiCGStab to converge again faster than CG and MR. The CPU-time improvement on the 32-node CM5 is nearly a factor of 1.5 compared to both competing methods.

As in the last section, we cross-checked the resulting difference norms and found the deviations being smaller than $10^{-6}$.

## 4 Summary and outlook

We have shown that the stabilized biconjugate gradient algorithm is a highly competitive tool for both the computation of the fermionic propagator in quenched lattice QCD and the calculation of the fermionic force in simulations of full QCD. BiCGStab turns out to be superior to conjugate gradient as well as minimal residual in particular in the region of small quark masses.

As to preconditioning, we only used standard even-odd splitting of the fermionic matrix. More sophisticated preconditioning methods are currently under investigation.

# Acknowledgements

Part of this work was supported by the Deutsche Forschungsgemeinschaft Grant Schi 257/3-2. Th. L. gratefully acknowledges S. Guesken and E. Laermann for useful discussions and hints.

# Figure Captions

Fig. 1: Iteration numbers $N$ in a quenched propagator computation as functions of $\kappa$, normalized to $N_{BiCGStab}$ on one given configuration, with $r = 10^{-12}$. The vertical lines enclose the quark mass range mentioned in the text.

Fig. 2: CPU-time improvement factors on the 32-node CM5 as functions of $\kappa$ normalized to $N_{BiCGStab}$ on one given configuration.

Fig. 3: $\log_{10} r$ as function of the number of iteration steps.

Fig. 4: Iteration numbers $N$ as functions of $\kappa$, normalized to $N_{BiCGStab}$ on a sample of ten statistically independent configurations for $r = 10^{-9}$.

Fig. 5: Iteration numbers $N$ as functions of $\kappa$, normalized to $N_{BiCGStab}$ on a sample of ten statistically independent configurations for $r = 10^{-12}$.

Fig. 6: CPU-time improvement factors on the 32-node CM5 as functions of $\kappa$ normalized to $N_{BiCGStab}$ on a sample of ten statistically independent configurations for $r = 10^{-12}$.

Fig. 7: Iteration numbers $N$ in a HMC simulation as functions of $\kappa$ normalized to $N_{BiCGStab}$ on a sample of ten configurations.

Fig. 8: Iteration numbers $N$ in a HMC simulation as functions of $\kappa$ normalized to $N_{BiCGStab}$ on a sample of ten configurations. The determinant is simulated on the even subspace.

Fig. 1

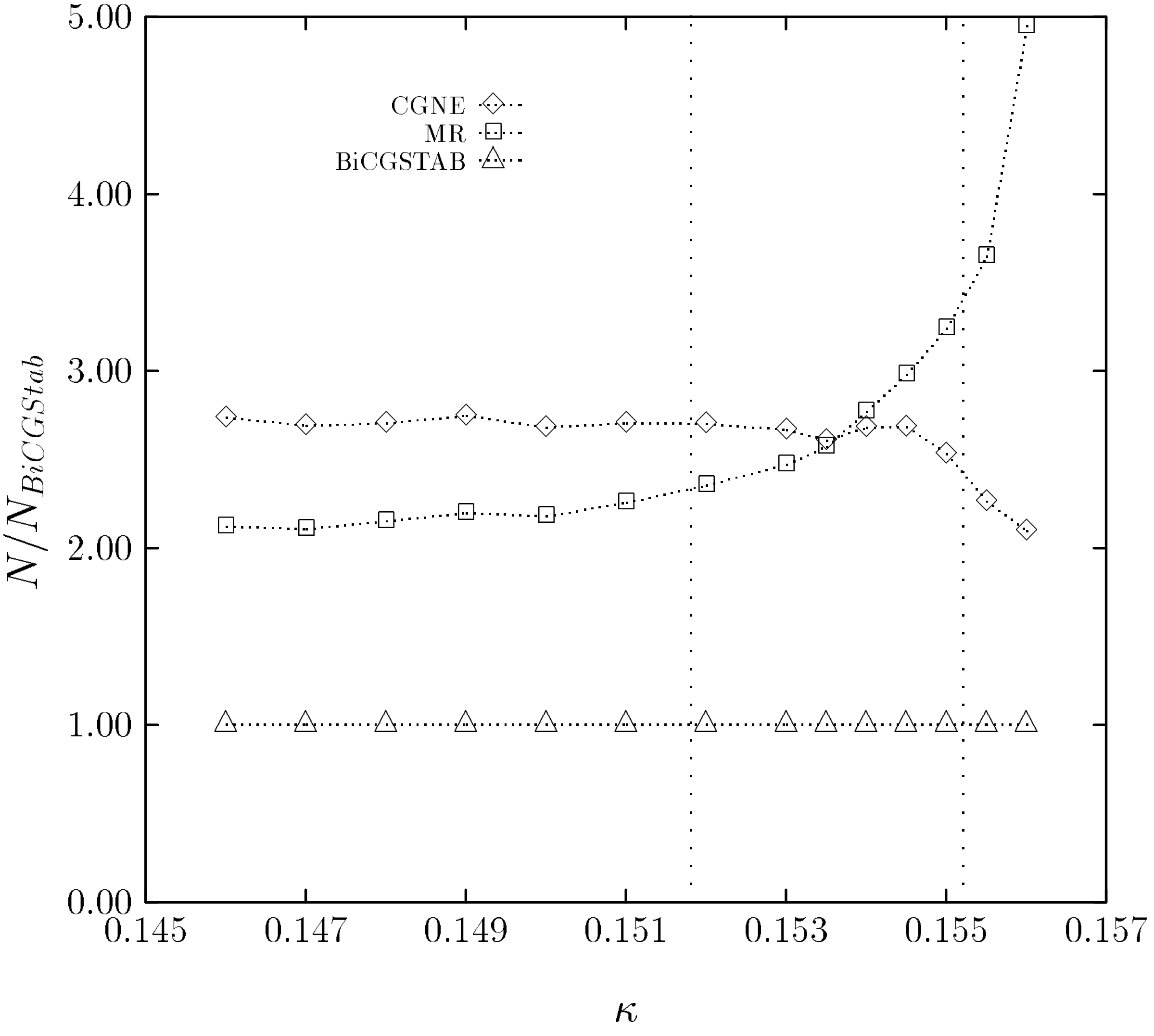

Fig. 2

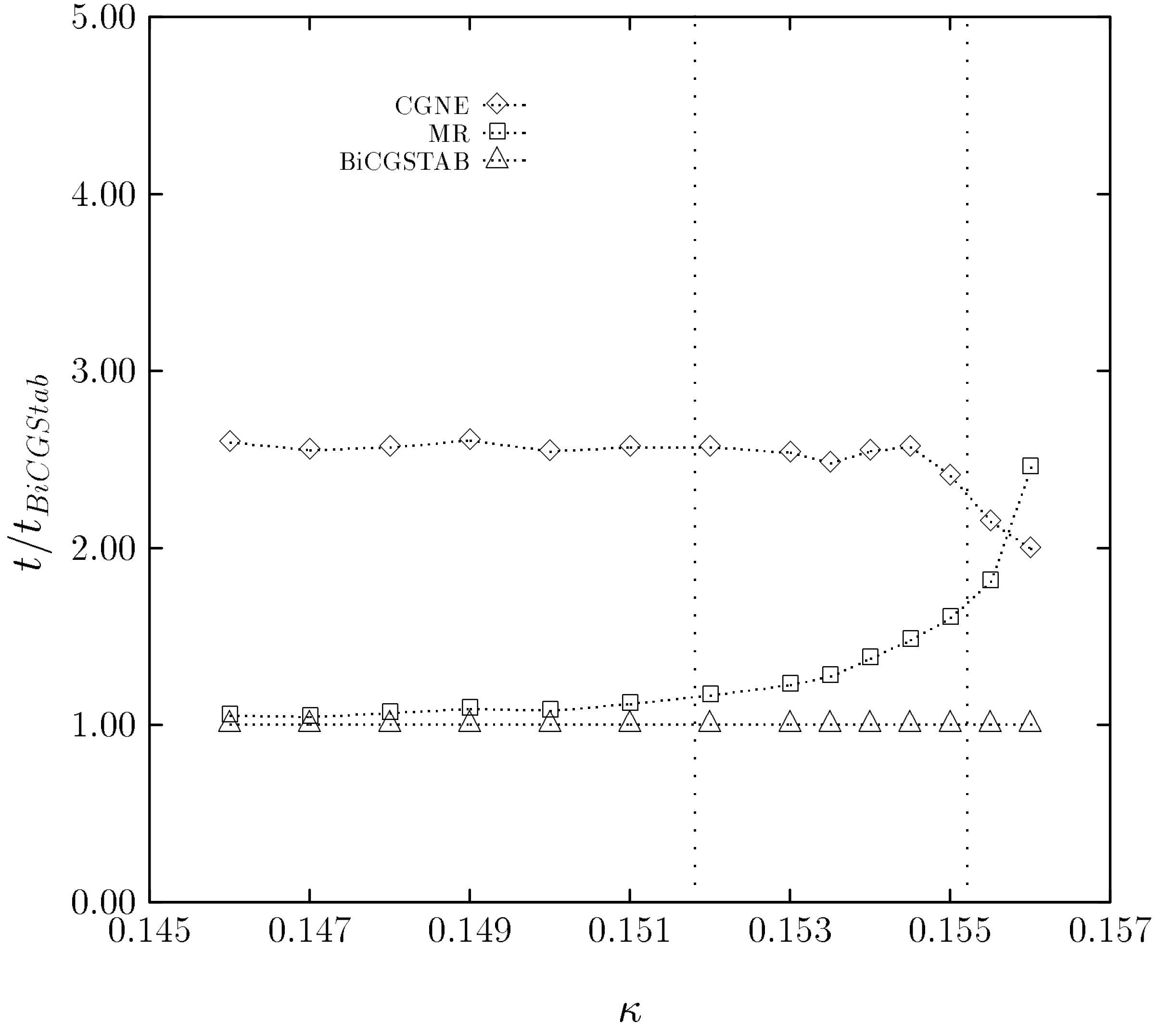

Fig. 3

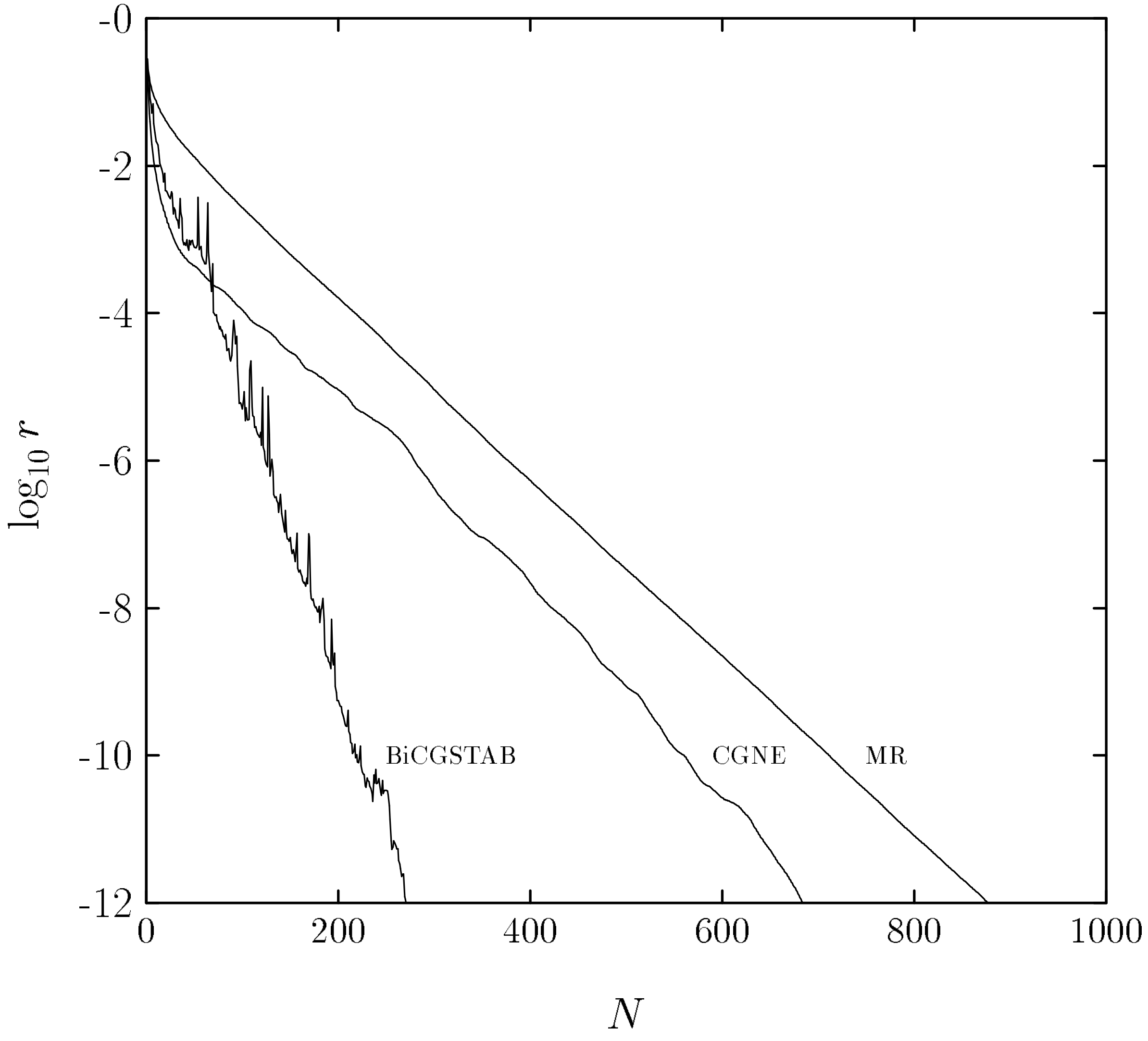

Fig. 4

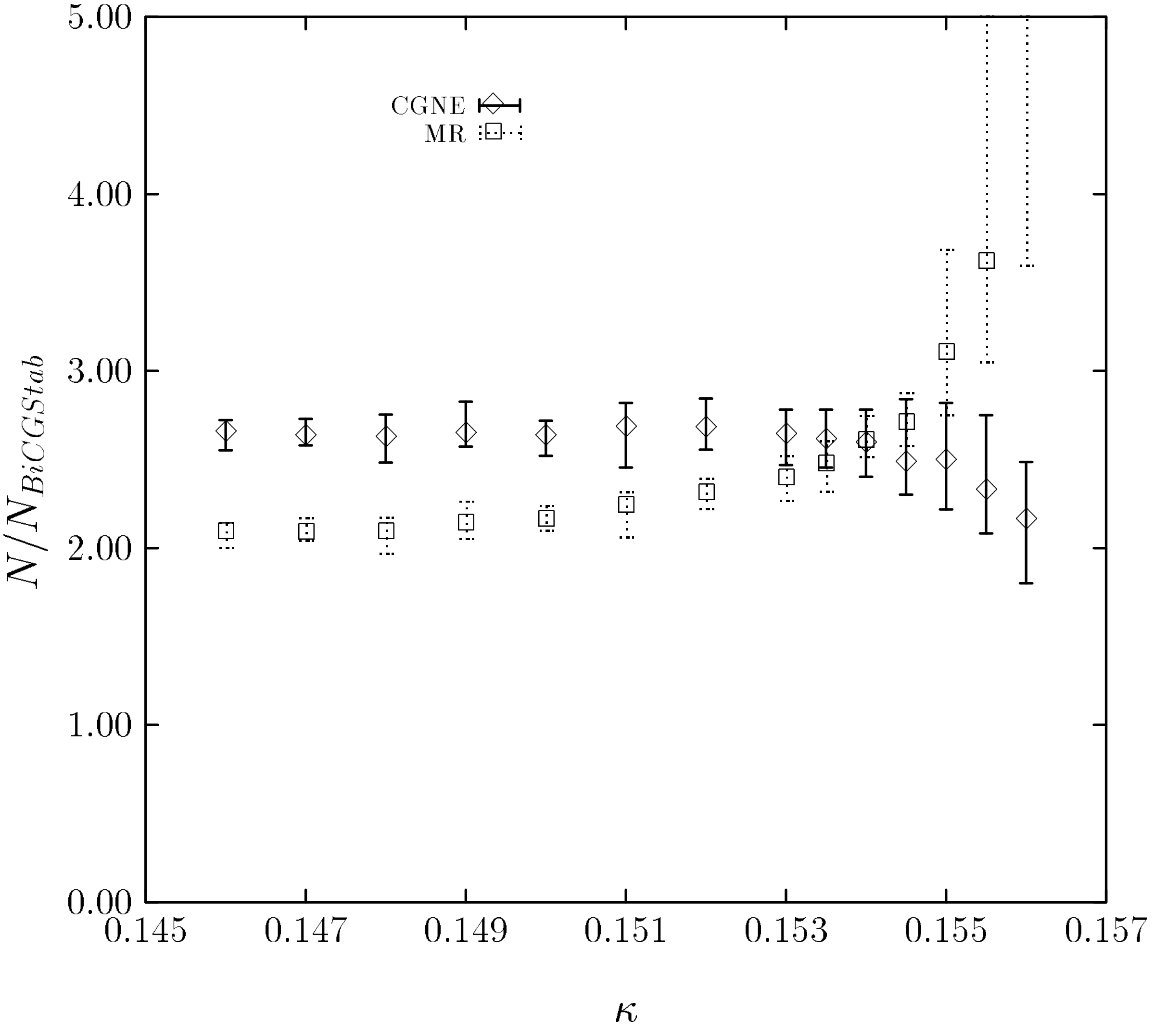

Fig. 5

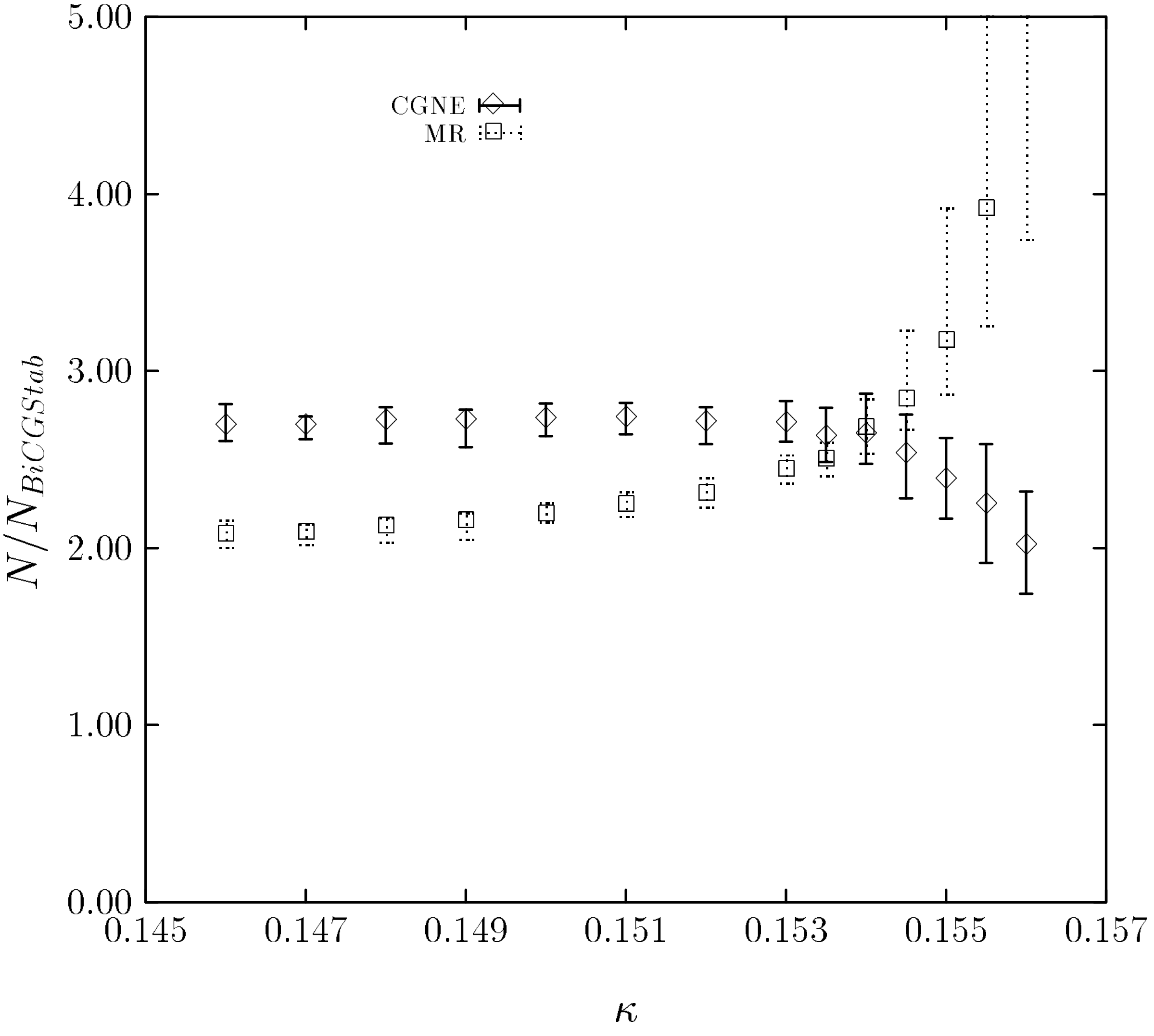

# Fig. 6

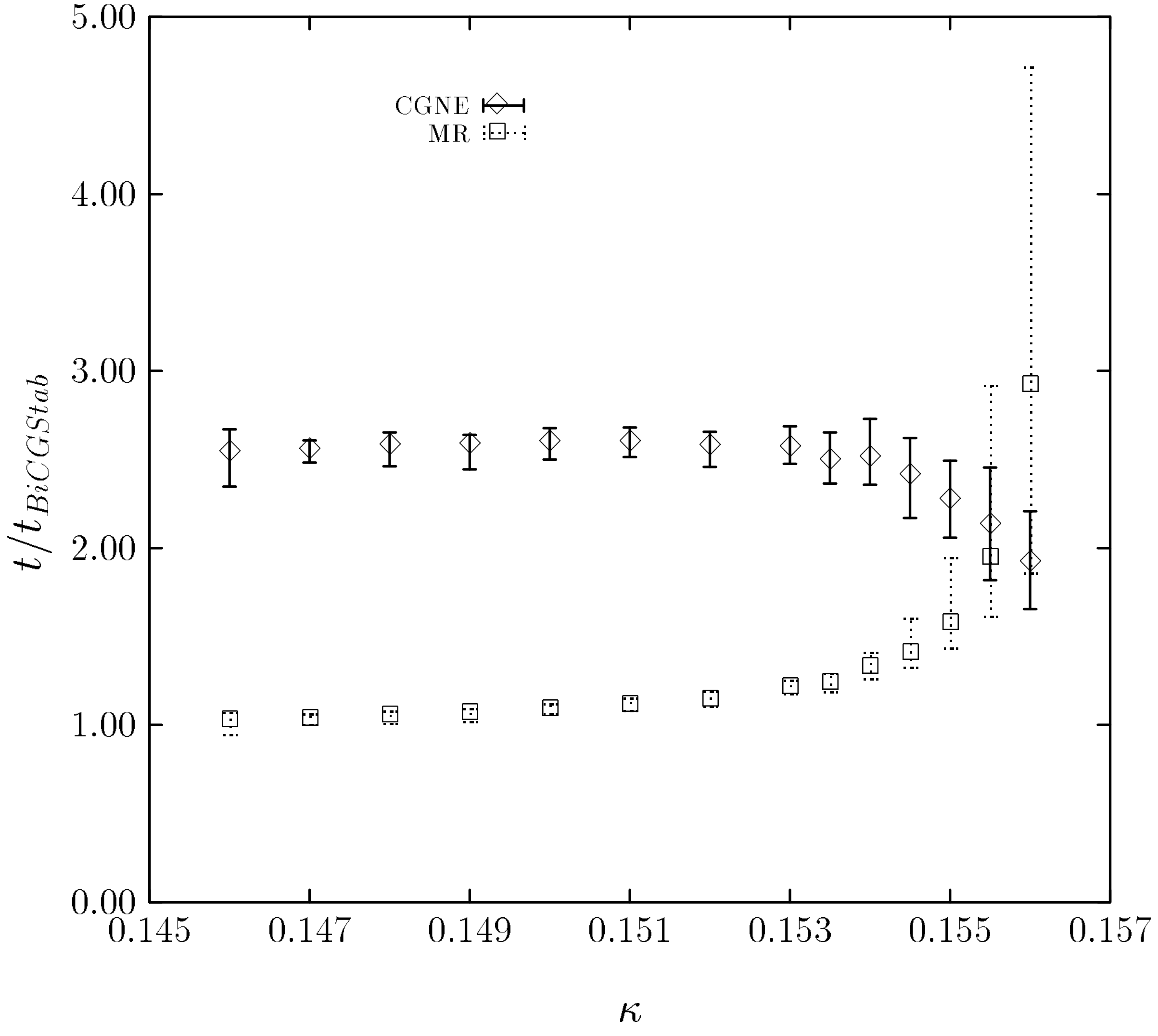

Fig. 7

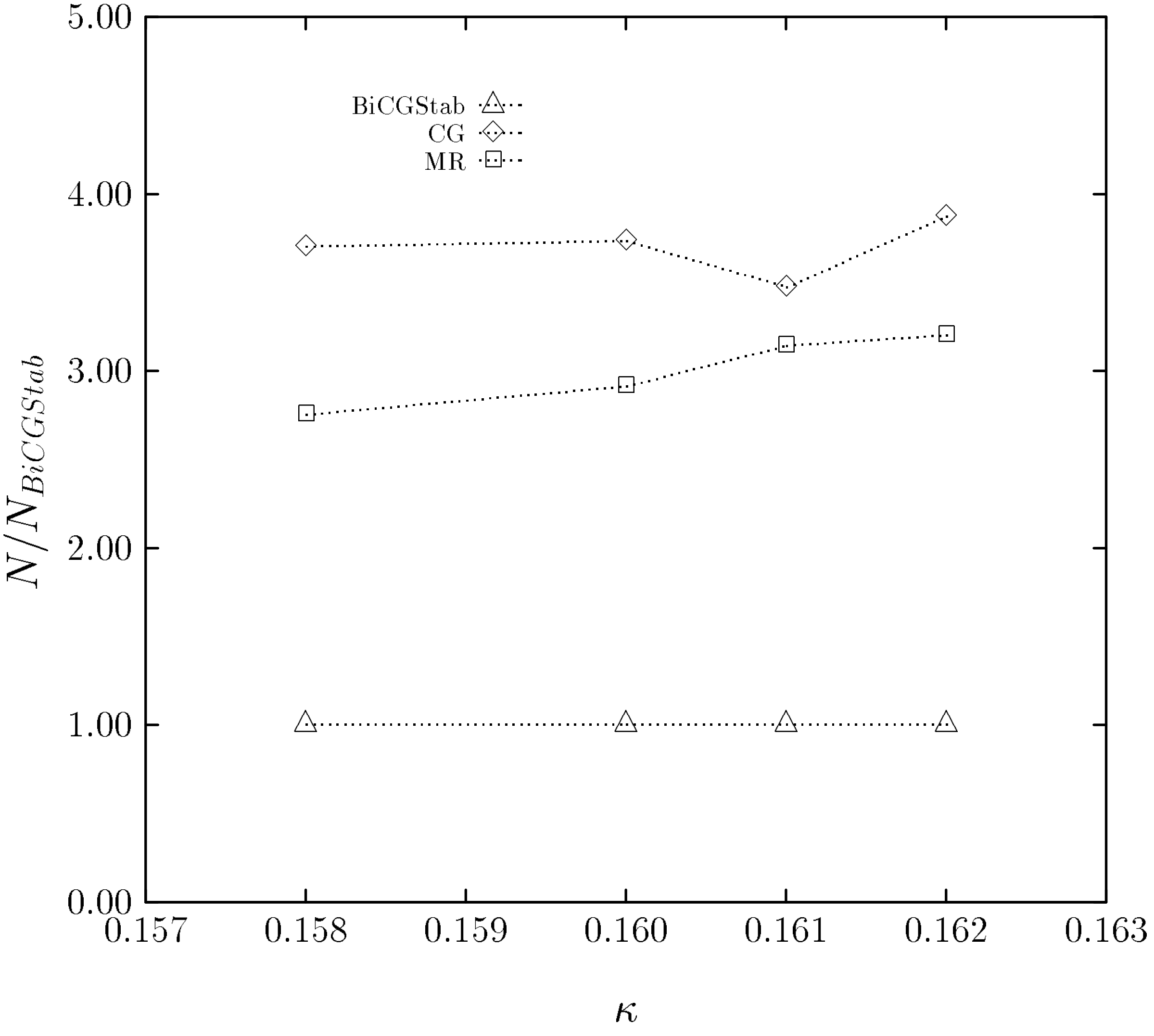

Fig. 8

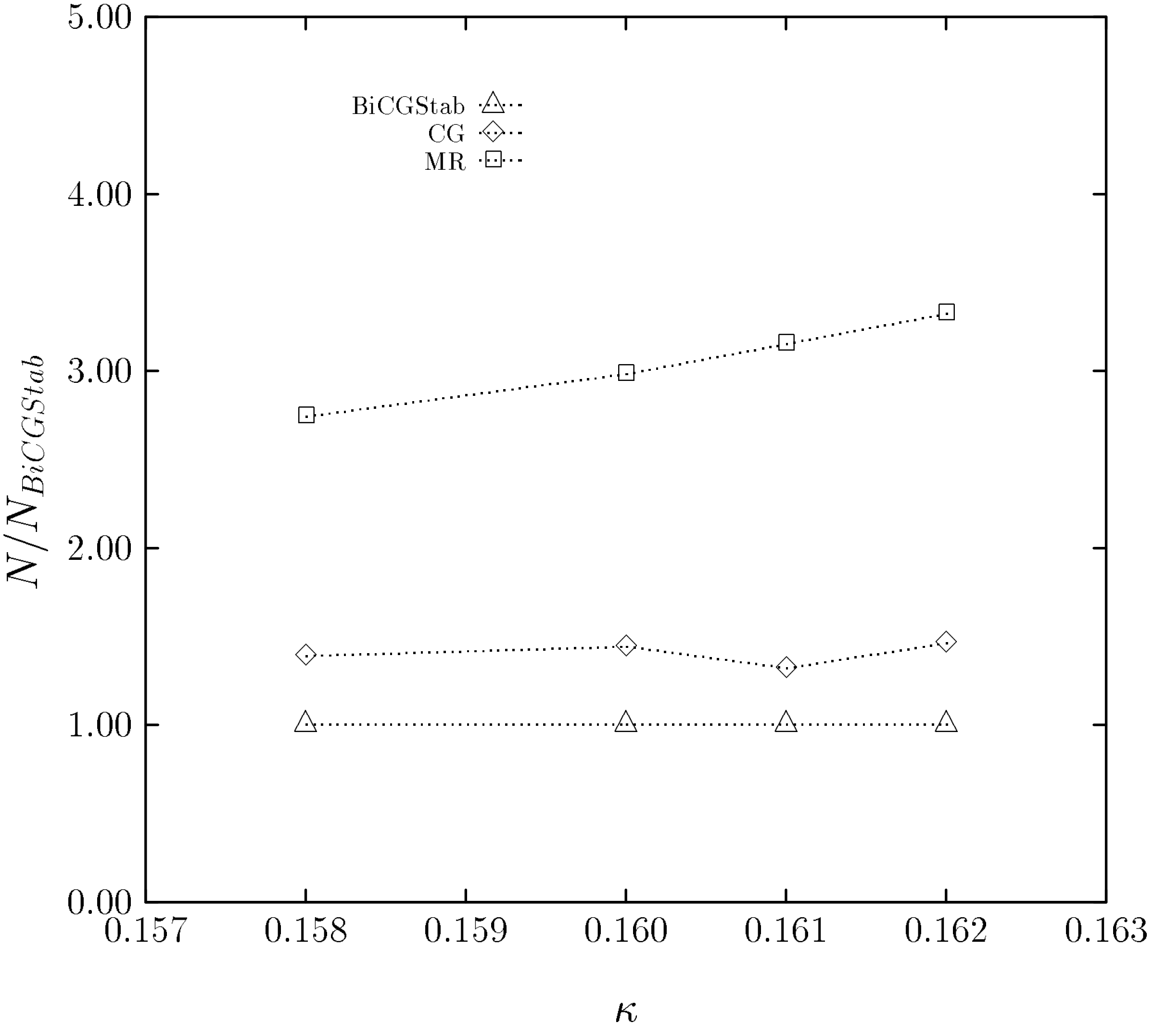